# Preventing COVID-59

Yaniv Erlich[1,*], Daniel Douek[2,*]

1 The Interdisciplinary Center (IDC) Herzelia, Herzelia, Israel
2 Vaccine Research Center, National Institute of Allergies and Infectious Diseases, NIH, Bethesda, Maryland

* To whom correspondence should be addressed (yaniv@eleventx.com; ddouek@mail.nih.gov)

**Abstract**
SARS-CoV-2 is the third betacoronavirus to enter the human population in the past 20 years, revealing a concerning pattern. Clearly, preventing a future pandemic from such viruses is a critical priority. Previous studies have shown that shRNAs can be powerful suppressors of RNA viruses in transgenic animals and substantially reduce transmission. Thus, we propose the introduction of anti-betacoronavirus shRNAs using CRISPR/CAS9 gene drive into the horseshoe bat population, the natural reservoir of those viruses, to combat this pandemic threat at its source. Importantly, our approach is not expected to create any harm to bats and can benefit other animals in the ecosystem that contract betacoronaviruses from bats. We map the ethical and the technical aspects and suggest guidelines for moving forward with this proposal.

**Main Text**
Coronaviruses are zoonotic viruses that have the potential to cause pandemics and to induce severe respiratory disease in humans. As of today, two alpha- and five beta-coronaviruses are known human pathogens: HCoV-NL63, HCoV-229E, HCoV-OC43, HCoV-HKU1, SARS-CoV, MERS-CoV, and SARS-CoV-2 [1]. The first four are endemic viruses that induce a relatively mild disease responsible for a proportion of common colds. The other three viruses can induce severe respiratory disease with potential long term effects and case fatality rates that range from ~35% for MERS [2] to approximately 0.5%-1% [3] for SARS-CoV-2, which might be higher with the new emerging variants [4]. As of today, no highly effective treatment exists for any of these viruses besides vaccination.

Alarmingly, the past 20 years have witnessed a repeated pattern in which pathogenic betacoronaviruses enter the human population and create severe diseases. The first case in this pattern is the 2002 SARS-CoV outbreak, which is likely the result of a zoonotic transmission from horseshoe bats (*Rhinolophus sinicus* and *Rhinolophus affinis*) in south-east China to humans via an intermediary host [5,6]. The second case is the 2012 MERS outbreak, in which the most likely chain of events was transmission from bats to humans via camels as intermediate hosts [7]. The specific bat species has yet to be identified but studies suggest that insectivorous bats in the Horn of Africa are the probable source. As for SARS-CoV-2, while a lab escape cannot be fully written off [8,9], a spillover from bats, perhaps via an intermediary species, is nevertheless the most acceptable scientific explanation [9–12] and the working hypothesis of this manuscript.

With this picture in mind, it is quite plausible that humanity will have to deal with repeated outbreaks of coronavirus infections in the foreseeable future, with bats being the most likely source of the virus. First, the contact surface with natural bat habitats may increase due to population growth and rapid urbanization, which could lead to increased risk of spillovers. The consequence is an increased likelihood of exposure to bat viruses. Second, immunity to such future viruses may not exist in the human population or, at least, may not be protective. Recent reports show that even a relatively small number of mutations in the Spike protein can reduce the effectiveness of infection- or vaccine-induced antibodies[14] and that antibodies that target the SARS-CoV spike generally do not neutralize the SARS-CoV-2 spike [15].

Thus, pandemic preparedness has become, with very good reason, a prime goal for the future. It encompasses the development not only of vaccines and therapeutics at a scientific level, but also the infrastructure to manufacture, test and deploy them, as well as worldwide immune surveillance in humans and pathogen surveillance in the wild. That task, while unquestionably necessary, is daunting and fraught with uncertainties. Time is a factor and this COVID-19 pandemic serves to underscore the urgency of the problem because it is not unreasonable to contend that a "SARS-CoV-3" pandemic may come sooner rather than later.

In the past few years, a number of studies have proposed that gene drives in reservoir animals could be a worth considering as a tool for combatting zoonotic pathogens [16]. In traditional population genetics, alleles are inherited to only 50% of the offspring. This means that introducing new neutral alleles in the wild type stationary population is a futile process that will result in their removal after a few generations. In gene drives, the alleles are introduced as part of a selfish genetic element that induces super-Mendelian inheritance, in which more than 50% of the progeny will receive the alleles. Thus, these alleles can rapidly monopolize the allelic pool within the target population and become fixed. Recent years have witnessed a series of breakthroughs in creating these gene drives by using CRISPR/Cas9 technology [17,18]. The locus of interest is equipped with the modified allele, together with a Cas9 and gRNA cassettes that induce double strand breaks in the locus on the homologous chromosome. The homologous recombination system uses the modified locus as a template to repair the breaks, effectively copying the allele to the other chromosome.

Previous studies have generally proposed two main strategies to control pathogens using gene drives [19]. The first strategy is to induce population suppression of the host species; for example, by reducing the proportion of fertile females until no females are available to produce progeny. The other strategy is to equip the host species with specific defense mechanisms against the pathogen in question. This could be achieved by altering a critical receptor entry point or adding a genetic element that confers resistance, such as a nanobody. Both types of gene drives have been proposed as mechanisms to control various host populations of human pathogens, including *Aedes aegypti* and *Anopheles gambiae* mosquitoes that are the primary vector of dengue and malaria, respectively [20,21]. With respect to mammals, studies have proposed the use of gene drives to control invasive rodent populations [22].

Taking our cue from previous work on gene drives, we suggest a similar strategy also for controlling betacoronavirus in bats. Specifically, we propose here that a future betacoronavirus outbreak could be mitigated using a bat gene drive of a short hairpin RNA (shRNA) element against betacoronavirus (**Figure 1**). shRNAs are small genetic elements that can target RNA viruses for destruction [23]. They are processed by a series of RNA interference (RNAi) enzymes and eventually a short (19-22nt) segment from the original shRNA, called the guide strand, is loaded into the Ago2 enzyme [24]. This enzyme scans long RNA molecules and if there is reverse complementation, it cuts the long RNA strand, which usually results in its destruction. Unlike humoral immune responses that are generally restricted to viral envelope proteins, any 19 to 22nt stretch in the viral genome could be a potential target for an shRNA, opening up many more possibilities for the targeting of conserved regions among beta-coronaviruses. Finally, cocktails of RNAi molecules can be used to prevent viral escape and increase the potency against multiple strains [25].

**Figure 1**

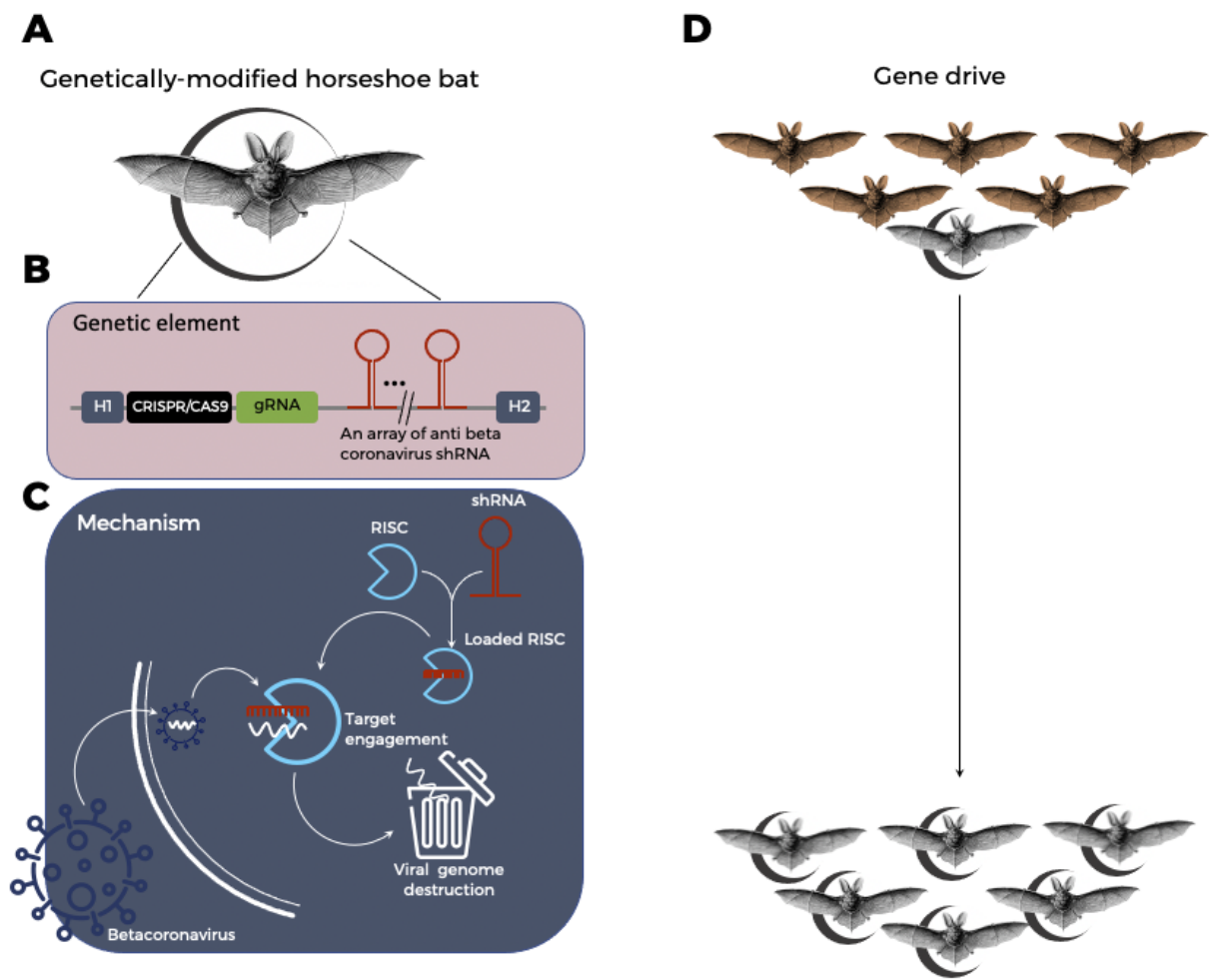

**Figure 1: Schematic representation of a gene drive to prevent betacoronavirus spillovers from bats (A)** The genetically modified bat **(B)** The genetic element consists of an array of shRNA triggers that target conserved regions across betacoronavirus. It also encodes for a CAS9 protein, gRNA, and homologous segments (H1 and H2) that create a homing endonuclease that replaces the wildtype allele in the homologous chromosome with the genetic element **(C)** the shRNAs mature and are loaded into the RISC complex. Upon betacoronavirus infection, the RISC complex engages with the RNA genome of the virus and destroys it **(D)** The modified bats are released in strategic locations of high horseshoe bat-human contacts. After multiple generations, the element spreads in the bat population and confers resistance against betacoronaviruses.

Importantly, previous studies have described transgenic stock animals with anti-viral shRNA activity. For example, Lyall et al. [26] reported the generation of transgenic chickens with shRNA against swine flu. Next, they challenged these transgenic chickens with highly pathogenic avian flu and placed them in the same habitat with naive chickens. As a control, they followed the same process but this time they challenged naive chickens instead of shRNA transgenic ones and placed them with naive chickens in another habitat. Interestingly, they found no transmission of the virus in the scenario where the challenged chickens carried the shRNA transgene compared to the latter scenario of challenging naive chickens. Similarly, another study reported transgenic pigs with shRNA against the foot and mouth disease virus (FMDV), which causes severe disease in naive piglets [27]. Challenging the transgenic pigs with FMDV resulted in substantial reduction of the disease phenotype compared to naive pigs. Finally, another study engineered transgenic pigs with shRNA against classical swine fever virus (CSFV). Interestingly, the immunity was transmitted to the next generation with the inheritance of the shRNA element, showing that this approach can propagate in populations [28]. The field has yet to develop a transgenic animal engineered with an shRNA targeting beta-coronavirus. However, a previous key study has examined the prophylactic activity of a nasally delivered cocktail of two siRNA against SARS-CoV in mice and rhesus macaques [29]. Both animal types that received the cocktail were largely immune to SARS-CoV and virus free in their upper respiratory tract, suggesting that they were not infectious. As shRNA transgenes are generally more effective than externally administered siRNAs, the results in rhesus macaques and mice suggest that it is quite likely that betacoronavirus can similarly be targeted with an shRNA transgene. Equally important, in none of the studies above did the animals show any signs of adverse effects due to the shRNA transgene or the siRNA treatment. Taken together, these studies strongly suggest that an anti-beta-coronavirus shRNA transgene could be safe and effective in wild animals.

In parallel to the advances in shRNA transgenes, the field has witnessed the advent of CRISPR-based gene drives. These gene drives are highly flexible and relatively effective. For mammals, so far there has been one report of a CRISPR-based gene drive in mice [30]. However, the efficiency was only 72% transmission (compared to over 95% in other non-mammalian studies) and the homing effect was limited to females. Nevertheless, this initial study clearly shows the feasibility of implementing such gene drive systems in mammals. We can expect that with further understanding and tuning of the genetic control and homing mechanisms in mammalian germlines, these systems will improve.

We posit that such gene drive is ethically sound and reasonable. From an environmental perspective, our approach aims neither to cause population collapse nor to change any endogenous genes in bat genomes. Rather, we propose to introduce a genetic element that is expected to have a neutral effect on the fitness of bats. Indeed, previous work has speculated that bats harbor high rates of viruses, as part of mutualistic symbiosis, for instance by forming a "biological weapon" that is non-harmful for the bats but can affect predators [31]. However, we are not aware of any evidence that supports this speculation. Even if such mutualism benefits the bats, a gene drive is likely to be far less harmful for bats than the current alternative of actively culling bats in certain regions due to fear of disease spreading [32]. Moreover, spillover of viruses from bats endangers other animals. For example, in 2016, an alphacoronavirus spillover from bats resulted with the outbreak of swine acute diarrhea syndrome (SADS) in Guangdong that eradicated over 24,000 piglets [33]. The Covid19 pandemic affected a variety of mammals, from dogs to lions [34]. Thus, our approach is not only beneficial for humans but also for other animals in the ecosystem.

However, our proposal should not be taken as a carte balance for gene drive in bats. We envision that such a project will be governed by an international committee that will include officials from relevant Government regulators, bioethicists, infectious disease experts, zoologists, and representatives from local communities. The tests for gene drives will first begin in the lab, then in specific locations in the field, where it can be controlled to some extent (e.g. small islands), then through a pilot phase in specific areas, and only then it will progress into a full effort to introduce the transgene to the population. In each step, there will be a clear set of exit criteria for the safety of the system and its efficacy. For example, one crucial question that needs to be addressed is whether eradicating beta coronaviruses from bats can make them more susceptible to even more harmful viruses, such as hendra virus. While examples of viral competitive exclusion are rather scarce [35], this question can be addressed in a controlled setting of challenging beta-coronavirus positive horseshoe bats with various viruses and checking their viral load compared to the beta-coronavirus free bats.

When considered in the context of the tremendous toll COVID-19 continues to have on humanity, a gene drive does not seem unreasonable. Over two million people have lost their lives directly from SARS-CoV-2 infection in ten months despite unprecedented measures to contain and control the pandemic. Per annum, this number is ten to twelve times the death toll due to malaria, which is considered the prime candidate for a gene drive approach. In the US alone, it has been estimated that at least 2.5 million person-years have been lost due to the pandemic by October 2020 [36]. The impact of illnesses due to betacoronaviruses stretches far beyond mortality counts. These viruses can have long lasting consequences for the health of affected individuals, including cardiac, pulmonary, neurological, and cognitive symptoms that may affect people for years [37]. The psychological effects due to the pandemic and the countermeasures have produced a wave of depressive mood disorders, domestic violence, and a range of pathological behaviours[38]. Finally, the economic toll of the beta-coronavirus infection is massive. The relatively small SARS-CoV outbreak wiped out about $50 billion from the global economy [39] and the economic damage of SARS-CoV-2 continues to grow as we write these lines. Future outbreaks, even if eventually contained, may severely affect the global economy. All of these factors strongly argue for taking strong steps to reduce the risk of such future pandemics.

Beyond the ethical aspects, our proposal is not an easy feat from a technical standpoint. Perhaps the most prominent challenge is the surprisingly long sexual maturation time of bats that is typically more than two years. That means that it will take a considerable amount of time to develop such gene drives in the laboratory and to generate sufficient impact on the bat population in the wild once implemented. However, we do not need to get the genetic element to get to every horseshoe bat in the world. Rather, we can focus on strategic habitats that exhibit frequent contacts with humans and exhibit the highest risk and introduce a series of shRNA elements to reduce the chance for escape mutants between the time of the release of the transgenic bats to the likely time of getting the shRNA alleles to sufficient levels. The long time it would take to execute this proposal should not deter us.

These obstacles may be mitigated with a stepwise approach. The first step is long term commitment from governments at a national and international level to secure funding for such a program. In this context, the issues of governance and global equity are paramount. Second, the program would require the establishment of substantial infrastructure to produce transgenic bats before their release into the wild in locations that constitute the main contact surface with humans. Introduction of a large number of bats will reduce the number of generations required to establish the transgene and achieve herd resistance.

Clearly, our approach is far from orthodox but it aims to eliminate the reservoir of these pandemics. Karl Marx wrote: "History repeats itself, first as tragedy, second as farce." With betacoronaviruses history has already repeated itself, first as a tragedy and now as a worse tragedy. We now have the components to control the emergence of these viruses with a proactive approach, and if successful, we envision that the same strategy could be tailored to combat other viruses whose main reservoir is bats, such as Ebola and Nipah, that are on the WHO watch list of pandemic threats. The world should not have to tolerate another pandemic caused by a human-adapted bat betacoronavirus but the clock is ticking.